\documentclass[aps,pre,twocolumn,floats]{revtex4}
\usepackage{epsfig}
\usepackage{bm}
\usepackage{latexsym}
\usepackage{ifthen}
\begin{document}

\newcommand{\hide}[1]{}
\newcommand{\tbox}[1]{\mbox{\tiny #1}}
\newcommand{\half}{\mbox{\small $\frac{1}{2}$}}
\newcommand{\sinc}{\mbox{sinc}}
\newcommand{\const}{\mbox{const}}
\newcommand{\trc}{\mbox{trace}}
\newcommand{\intt}{\int\!\!\!\!\int }
\newcommand{\ointt}{\int\!\!\!\!\int\!\!\!\!\!\circ\ }
\newcommand{\eexp}{\mbox{e}^}
\newcommand{\bra}{\left\langle}
\newcommand{\ket}{\right\rangle}
\newcommand{\EPS} {\mbox{\LARGE $\epsilon$}}
\newcommand{\ttimes} {\mbox{\tiny \ $^{\times}$ \ }}
\newcommand{\ar}{\mathsf r}
\newcommand{\im}{\mbox{Im}}
\newcommand{\re}{\mbox{Re}}
\newcommand{\bmsf}[1]{\bm{\mathsf{#1}}} 

\newcommand{\be}[1]{\begin{eqnarray}\ifthenelse{#1=-1}{\nonumber}{\ifthenelse{#1=0}{}{\label{e#1}}}}
\newcommand{\ee}{\end{eqnarray}}


\title{Quantum pumping: The charge transported due to a translation of a scatterer}

\author{Doron Cohen$^{1}$, Tsampikos Kottos$^{2}$ and Holger Schanz$^{2}$}

\affiliation{
$^{1}$ Department of Physics, Ben-Gurion University, Beer-Sheva 84105, Israel \\
$^{2}$ Max-Planck-Institut f\"ur Dynamik und Selbstorganisation, \\
Bunsenstra\ss e 10, D-37073 G\"ottingen, Germany
}


\begin{abstract}
The amount of charge which is pushed by a moving scatterer is $dQ = -G dX$,
where $dX$ is the displacement of the scatterer.  The question is what is
$G$.  Does it depend on the transmission $g_0$ of the scatterer?  Does the
answer depend on whether the system is open (with leads attached to
reservoirs) or closed?  In the latter case: what are the implications of
having ``quantum chaos" and/or coupling to the environment?  The answers
to these questions illuminate some fundamental aspects of the theory
of quantum pumping.  For the analysis we take a network (graph) as a model
system, and use the Kubo formula approach.
\end{abstract}

\maketitle


Consider an Aharonov-Bohm ring with a Fermi sea of non-interacting spinless
electrons as in Fig.1b.  Assume that the ring is either disordered or chaotic,
and that the temperature is known.  In such a case the ring has a well defined
Ohmic conductance $G^{\tbox{ohm}}$. This means that if we change the magnetic flux
$\Phi$ through the ring, then we have the electro-motive-force $-\dot{\Phi}$
and the current is ${\cal I} = G^{\tbox{ohm}} \times (-\dot{\Phi})$.  Therefore the
charge which is transported is $dQ=-G^{\tbox{ohm}}d\Phi$.  Next we can ask what is
the charge that is transported if we vary some other parameter ($X$) that
controls the potential in which the electrons are held (in practice it may be
a gate voltage).  Then, in complete analogy with Ohm's law, we expect to get
$dQ=-GdX$, where $G$ is a {\em generalized conductance}.  The calculation of
$G$ is the so-called problem of ``quantum pumping".

Most of the studies of quantum pumping were (so far)
about open systems (Fig.1e).
Inspired by Landauer who pointed out that $G^{\tbox{ohm}}$
is essentially the transmission of the device,
B{\"u}ttiker, Pretre and Thomas (BPT) have
developed a formula that allows the calculation of $G$ using
the $S$ matrix of the scattering region \cite{BPT,AvronSnow}.
It turns out that the non-trivial extension of this approach
to closed systems involves quite restrictive assumptions \cite{MoBu}.
Thus the case of pumping in closed systems has been left un-explored,
except to some past works on adiabatic transport \cite{BeRo,AvronNet}.
Yet another approach to quantum pumping is to use
the powerful {\em Kubo~formalism} \cite{pmc,pmo}.
In this Letter we report the first non-trivial demonstration
of this formalism. Namely, we illuminate the interplay
of ``quantum chaos" with non-adiabaticity and
environmental effects, and in particular we derive
specific results for the (generalized) conductance
in closed (chaotic) system.

        
To be specific we ask what is the amount of charge which is transported if we
make a displacement $dX$ of some scatterer or of some wall element
(``piston").  The answer to this question in case of the 1D system of Fig.1e
is well known.  Using the BPT formula one gets
\be{1} 
G \ = - (1-g_0) \times \frac{e}{\pi}k_{\tbox{F}}
\ee
where $k_F=\mathsf{m}v_F/\hbar$ is the Fermi wave number, 
and $g_0$ is the transmission of the scatterer. 
This result is analogous to the Landauer formula $G^{\tbox{ohm}}=(e^2/\hbar)g_0$.
The charge transport mechanism which is represented  
by Eq.(\ref{e1}) has a very simple heuristic explanation, 
which is reflected in the term ``snow plow dynamics" \cite{AvronSnow}. 
On the other hand in case of Fig.1a or Fig.1c, if we translate 
the scatterer at some constant velocity $\dot{X}$, it is clear that  
\be{2} 
G \ = - 1 \times \frac{e}{\pi}k_{\tbox{F}}
\ee
In Eq.(\ref{e2}) the transmission of the scatterer is assumed to be $g_0<1$
(in Fig.1a the role of $g_0$ is played by the relative size of the scatterer).
Irrespective of $g_0$ the steady state is a distribution that moves with the
same velocity as the scatterer.  (In the moving frame of the scatterer the
steady state is a standing wave).  Though $g_0$ does not influence the value
of $G$, it should be remembered that in the limit $g_0 \rightarrow 1$ it takes
an infinite time to get into the steady state.

What about Fig.1b? Here the system has an inherent time scale $\tau_{\rm cl}$ 
which governs the relaxation to an ergodic distribution in the laboratory frame
irrespective of the driving. The scatterer should push its way through 
the ergodizing distribution, and therefore its relative size (or its transmission) matters. 
Thus we expect to find $G\propto (1-g_0)$
in the analogous network model of Fig.1d. 
Moreover, we expect to have a dependence of $G$ 
on the overall transmission $g_T$ of the ring.  
Using the Kubo formalism we are going to derive
\be{3}
G = -\frac{e}{\pi}k_{\tbox{F}} 
\left[ \frac{1-g_0}{g_0}\right]
\left[ \frac{g_T}{1-g_T}\right]
\ee
and to discuss the conditions for its validity. 
In particular we are going to discuss whether
the effect of non-adiabaticity is to induce 
a crossover from Eq.(\ref{e3}) to Eq.(\ref{e1}).  
We are going to present both semiclassical 
and quantum mechanical derivations, 
and to consider the generalization 
of ``universal conductance fluctuations".

Our model system \cite{kottos} is the network of Fig.1d.  
It is composed of 1D wires (``bonds").  
The scatterer is represented by a delta function which is located
on some selected bond.  Thus the Hamiltonian ${\cal H}$ incorporates a
potential $U(x) = \lambda(\hbar^2/2\mathsf{m})\,\delta(x-X)$, 
where $X$ is the location of the scatterer along the bond. 
The parameter $\lambda$ determines the
transmission of the scatterer $g_0=(1+(\lambda/2k_{F})^2)^{-1}$.

\begin{figure}
\epsfig{figure=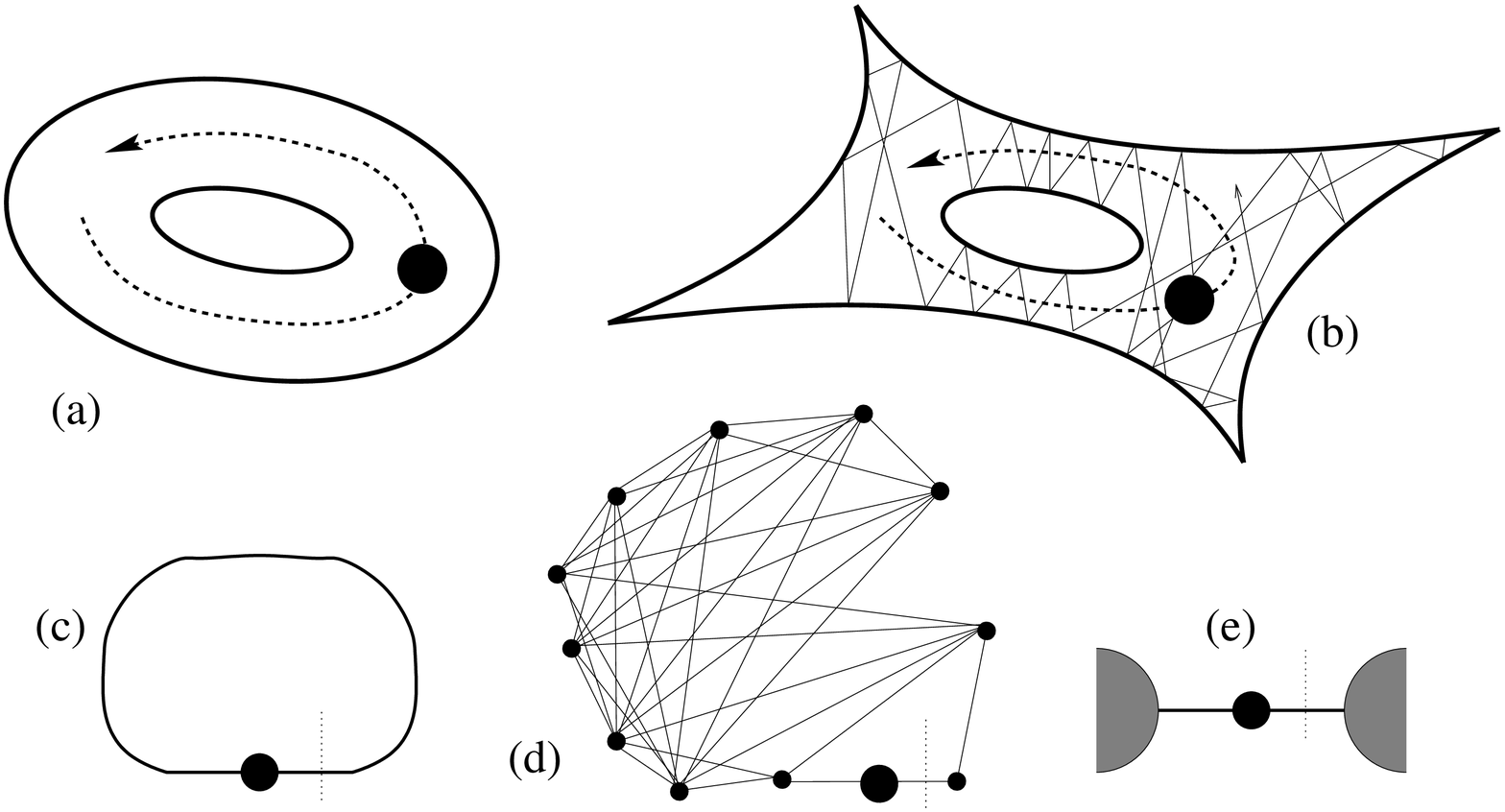,width=\hsize}
\caption{A scatterer (represented by a black circle) is 
translated through a systems that has a Fermi occupation of spineless non-interacting 
electrons. In (a) the system is a simple ring. 
In (b) it is a chaotic ring (Sinai billiard). In (c) and in (d) we have 
network systems that are of the same type of (a) and (b) respectively. 
In the network, the scatterer (``piston") 
is a delta function (represented as a big circle) located 
at $x=X$. The current is measure at a section 
(represented by a dotted vertical line) through $x=x_1$.
In (e) we have an open geometry with left and right leads that 
are attached to reservoirs that have the same chemical potential.} 
\end{figure}


The generalized conductance $G$ is in fact an element of the
generalized conductance matrix, that gives the response 
of the current ${\cal I}$ to the driving via $X$.  
With $X$ we associate the operator 
${\cal F}=-(\partial {\cal H} / \partial X) = \lambda(\hbar^2/2\mathsf{m})\,\delta'(x-X)$.  
Since $dX$ is a displacement, it follows that ${\cal F}$ is a Newtonian force.
We measure the current through another 
section $x=x_1$ along the same bond (see dotted line in Fig.1d):
\be{4}
{\cal I} = \frac{e}{2\mathsf{m}} \left( \delta(x-x_1)p + p\delta(x-x_1) \right)
\ee
The Kubo expression for $G$ can be expressed using 
a generalized Fluctuation-dissipation (FD) relation \cite{pmo}:
\be{5}
G \ = \ g(E_F) \int_0^{\infty}C_{E_F}(\tau)d\tau 
\ee
where $C_E(\tau)$ is the cross correlation function of ${\cal I}(\tau)$ and 
${\cal F}(0)$, and $g(E)$ is the one-particle density of states at the Fermi energy. 
Eq.(\ref{e5}) is the Kubo analog of the BPT formula. In fact the latter can be obtained 
as a special limit of the former \cite{pmo}. Eq.(\ref{e5}) would become the standard FD 
relation if we were looking for $G^{\tbox{ohm}}$. In such case the symmetric 
current-current correlation function would be involved, and consequently one 
would obtain $G^{\tbox{ohm}} = \half g(E_F) \tilde{C}_{E_F}(\omega{\sim}0)$. 
But in the case of our pumping calculation $C_E(\tau)$ is 
antisymmetric, and therefore such simplification is not possible.

Disregarding the assumption of having Fermi occupation, Eq.(\ref{e5}) has a classical 
derivation.  In such case $C_E(\tau)$ is the classical cross correlation function, 
and there is an implicit thermal averaging over $E_F$. The analogous quantum mechanical (QM)
derivation is based on perturbation theory. Later we are going to discuss its limitations.  
In the QM context $C_E(\tau)$ is the {\em symmetrized} correlation function. 
Furthermore, considerations that are based on a Markovian treatment of perturbation 
theory to infinite order, suggest the replacement
\be{6}
C_E(\tau) \ \ \mapsto  \ \   
C_E(\tau) \eexp{-(\Gamma/2\hbar) \tau} 
\ee
where $\Gamma \propto |\dot{X}|^{2/3}$ is related to the non-adiabaticity 
of the driving \cite{pmc}, or more generally it may incorporate also the 
influence of an external bath. Thus we get
\be{7}
&& 
G =  
\hbar\,g(E) \int_{-\infty}^{\infty} 
\frac{-i\tilde{C}_{E_F}(\omega)}{\hbar\omega-i(\Gamma/2)} 
\ \frac{d\omega}{2\pi}
\ee
where 
\be{0}
\tilde C_E(\omega)={\hbar\over 2g(E)}[C(E{+}\hbar\omega,E)
{+} C(E{-}\hbar\omega,E)]
\ee
and 
\be{-1}
&& C(E',E) = 
2\pi \sum_{nm} 
{\cal I}_{nm} \delta(E'{-}E_m) 
{\cal F}_{mn} \delta(E{-}E_n) 
\\
&& 
\ \ \ \ \ \ \ \ =
\frac{2}{\pi} 
\ \trc\left[ 
{\cal I} \ \im[{\mathsf G}(E')] \ {\cal F} \ \im[{\mathsf G}(E)] 
\right]
\label{e8}
\ee
In the first equality $C(E',E)$ is expressed 
using the eigen-energies $E_n$ and the matrix elements 
of ${\cal I}$ and ${\cal F}$. In the second equality 
it is expressed using 
$\im [{\mathsf G}] = -i\half({\mathsf G}^{+}{-}{\mathsf G}^{-})$
where ${\mathsf G}^{\pm}=1/(E{-}{\cal H}{\pm}i0)$ 
are the Green functions of the system.

Thus we have three options for calculation. 
In the classical treatment it is simplest to 
calculate directly the correlation function 
in the time domain. In the QM case we can 
use expressions for the Green function in 
order to make the calculation. Optionally 
we can express the conductance using the 
eigen-energies and the matrix elements of the 
associated operators:       
\be{9}
G \ = \ 
2\hbar \sum_n f(E_n)
\sum_{m(\ne n)}
\frac{\im\left[{\cal I}_{nm}{\cal F}_{mn}\right]}
{(E_m{-}E_n)^2+(\Gamma/2)^2}
\ee
where $f(E)$ is the Fermi occupation function. 
The latter expression is valid 
also in the strict adiabatic limit $\Gamma \rightarrow 0$ 
where it can be regarded as geometric magnetism \cite{BeRo}.


We can get a semiclassical estimate for $G$ 
by studying the classical correlation function $C(\tau)$.
But first we should define what 
classical calculation means 
in the context of this network model.  
Recall that $dX$ is displacement,
so ${\cal F}$ has the meaning of Newtonian force.
Therefore in the classical calculation ${\cal F}(t)$ 
consists of spikes whose {\em area} has
the meaning of impact ($=2\mathsf{m}v_F$).
Possibly it is more intuitive 
to think of the scatterer as
a rectangular barrier with two vertical walls.
The vertical walls of the scatterer
are regularized by giving them finite slops.
In such case the spikes of ${\cal F}(t)$ 
become short rectangular pulses of some 
duration $\tau_0$ and height $(2\mathsf{m}v_F)/\tau_0$. 
Obviously this regularization 
drops out from the final result, 
because the product ${\cal I}(t) {\cal F}(0)$ 
is weighted by the probability of having 
non zero ${\cal F}(0)$, which is $(v_F\tau_0)/(2L)$.   
The possibility to tunnel through
the scatterer is taken into account by
adopting a stochastic point of view.  
Namely, upon collision there is 
a probability $g_0$ to go through the scatterer 
(in such case there is no impact).
Using the above stochastic picture 
one deduces that the short time correlations are 
\be{10}
C(\tau) = 
\frac{v_{\tbox{F}}}{2L} e (2\mathsf{m}v_{\tbox{F}})
\left[(1-g_0) \sum_{\pm} \pm \delta(\tau\pm\tau_1) \right]
\ee
where $\tau_1=(x_1{-}X)/v_F$. However there 
are tails due to multiple reflections,   
leading (after geometric summation) to 
\be{11}
\int_0^{\infty} C(\tau) d\tau = 
-e \frac{\mathsf{m}v_{\tbox{F}}^2}{L} 
\left[ \frac{1-g_0}{g_0}\right]
\left[ \frac{g_T}{1-g_T}\right]
\ee
Substitution into Eq.(\ref{e5}) leads to Eq.(\ref{e3}).
The validity of the final result can be double checked 
by solving classical Master equation to find the  
(quasi) steady state solutions of the problem. 
The current in the steady state, to linear order 
in the rate of the driving, leads to the same 
result (Eq.(\ref{e3})) for the conductance.


We turn now to the proper quantum mechanical 
calculation. The matrix elements between eigenstates 
of the network are:
\be{12}
{\cal I}_{nm} &=& -i \frac{e\hbar}{2\mathsf{m}} 
\left(\psi^n\partial\psi^m - \partial\psi^n\psi^m\right)_{x{=}x_1} \\
{\cal F}_{nm} &=& -\lambda{\hbar^2\over 2\mathsf{m}}
\left(\psi^n\partial\psi^m + \partial\psi^n\psi^m\right)_{x{=}X}
\ee
where the gradient $\partial\psi$ should be interpreted 
as the average value of the left and right slopes. 
[To derive this result it is convenient 
to regard the delta function 
as a narrow rectangular barrier.] 
Without loss of generality we set 
from now on $X=0$.
It is convenient to express ${\cal F}_{nm}$  
using the wavefunction at $x=+0$. Thus we get
\be{14}
{\cal F}_{nm} = 
-\lambda{\hbar^2\over 2\mathsf{m}}
\left(\psi^n\partial\psi^m + \partial\psi^n\psi^m
-\lambda\psi^n\psi^m \right)_{x{=}+0}
\ee
Substitution of (\ref{e12}) and (\ref{e14}) into (\ref{e9}) 
leads to an expression that can be written in terms of the Green
function~$\mathsf{G}$. [An alternate procedure is to substitute 
in Eq.(\ref{e8}) the implied differential representation of the operators]:  
\be{15} 
G=-{\lambda e\over \pi}
\left({\hbar^2\over 2\mathsf{m}}\right)^2
\langle
\mathsf{G}^{R} \mathsf{G}^{I}_{xx'}
-\mathsf{G}^{R}_{xx'} \mathsf{G}^{I} 
+\mathsf{G}^{R}_{x} \mathsf{G}^{I}_{x'}
\nonumber \\
-\mathsf{G}^{R}_{x'} \mathsf{G}^{I}_{x} 
+\lambda( \mathsf{G}^{R}_{x'} \mathsf{G}^{I} 
-\mathsf{G}^{R} \mathsf{G}^{I}_{x'}) 
\rangle_{E} 
\ee
where $\langle\cdot\rangle_{E}=-\int dE\,f'(E)$ implies 
thermal averaging and we have defined 
$G^{R}=\re\,\mathsf{G}(x,x';E+i\Gamma/2)$,
and $G^{I}=\im\, \mathsf{G}(x,x';E+i0)$.  
The subscripts indicate derivatives with 
respect to $x$ and $x'$. The expression 
is evaluated for $x=+0$ and $x'=x_1$.

In case of a network the Green function is given by \cite{paths} 
\be{17}
\mathsf{G}(x,x';E)= - \frac{i}{\hbar v_F}\sum_p A_p \eexp{ik_E L_p}
\ee
where the sum extends over all the paths that start at $x$ and end at $x'$. 
The $A_p$ are the product of the associated transmission and reflection 
amplitudes 
($\mathsf{\scriptstyle T}_i$ 
or $\mathsf{\scriptstyle R}_i$ 
for each encountered vertex $i$), 
while $L_p$ is the total length of the path.  
Upon substitution in (\ref{e15}) we get a double sum $\sum_{pq}$ 
over paths with endpoints $x=+0$ and $x'=X_{1}$. In a term that
involves derivatives the amplitude $A_p$ ($A_{q}$) is multiplied 
by a sign factor $s$ and/or $s'$, which indicates respectively 
the initial and final sign of the velocity.
Gathering all the contributions, one ends up with 
\be{-1}
G = -\frac{e}{\pi}k_F
\sqrt{\frac{1{-}g_0}{g_0}} \sum_L \sum_{p,q\in L}
s_p' s_{q} 
\,\im \left[ \frac{1}{\mathsf{\scriptstyle T}_0} A_p A_{q}^{*} \right] 
\eexp{-L/L_{\Gamma}}
\ee
where $L_{\Gamma}=2\hbar v_F/\Gamma$.  Above we neglected off diagonal terms
that involve pairs of trajectories with different lengths $L$. This is
justified if the energy averaging is over a sufficiently large range.  
It is also important to realize that any trajectory that starts at $x=+0$ and
departs in the positive direction represents in fact the contribution of two
degenerate {\em paths}: one starts with a positive velocity, while the other
starts with a negative velocity but is immediately reflected. Assuming that 
this is the only significant length degeneracy we get
\be{18}
G = -\frac{e}{\pi}k_F
\frac{1-g_0}{g_0} \sum_{p} 
s_p s_p' |A_p|^2 
\eexp{-L_p/L_{\Gamma}}
\ee
The sum is over the same paths as in Eq.(\ref{e17}), 
and it can be verified that the above mentioned  
degenerate paths adds correctly. 
The summation over $p$ involves a geometric sum 
in $[g_T-(1-g_T)]$ and gives the factor $g_T/(1-g_T)$. 
Thus we see that a careful treatment within the framework 
of the diagonal approximation recovers the classical result.


We turn to discuss the validity of our result.  
The derivation of the Kubo formula Eq.(\ref{e9}) 
assumes $\Gamma \ll \Delta_b$.
By definition the bandwidth 
$\Delta_b \approx \hbar/\tau_{cl}$ is 
the energy range $|E_n-E_m|<\Delta_b$ 
for which ${\cal I}_{nm}{\cal F}_{mn}$  
are non negligible. It is determined  
by the classical correlation time $\tau_{cl}$ 
that characterizes $C(\tau)$.    
For a generic chaotic system the mean level 
spacing is $\Delta \propto \hbar^{d}$,  
where $d$ is the dimensionality of the system.
Hence the bandwidth in dimensionless 
units is $b=\Delta_b/\Delta \propto \hbar^{1-d}$.
It follows that $b \gg 1$ is the generic 
case for any quantized chaotic system. 
For a chaotic network $d=1$, and $b$ 
is roughly equal to the number of bonds.

There is a practical implication of the 
the above discussion to pumping in general. 
For some geometries we have $\tau_{cl} \ll \tau_{1}$. 
The notable example is the dot-wire geometry 
of Ref.\cite{pmo} where $\tau_{cl}$  
is related to the motion inside the dot, 
while the current is measured outside 
at a section on the (very long) wire.  
Thus we may have $\Gamma > \hbar/\tau_1$ in Eq.(\ref{e6}) 
without breaking the validity condition 
$\Gamma < \hbar/\tau_{cl}$. Consequently we stay only 
with the short time correlations in Eq.(\ref{e10}),  
and we get Eq.(\ref{e1}) rather that  Eq.(\ref{e3}).

An additional assumption enters into the derivation of the generalized FD relation and 
the associated Green function expression. Namely, it is assumed that $\Gamma \gg \Delta$. 
In order to test the significance of this assumption, we consider a generic network with 
$b =35$. In Fig.2 we plot the exact result for $G$ as a function of $\Gamma$. We also plot 
the dispersion of $G$. One may regard this dispersion as the analog of {\em universal 
conductance fluctuations}. As $\Gamma$ becomes larger than $\Delta$ these fluctuations are 
smoothed away. 
We also see that the result is quite 
insensitive to the exact value of $\Gamma$ 
as long as \mbox{$\Gamma \ll \Delta_b$}, 
which is the regime where the quantum mechanical 
derivation of the Kubo formula makes sense. 
Throughout all regimes the diagonal approximation is very precise, as
expected for quantities related to the statistics of matrix elements.

\begin{figure}
\epsfig{figure=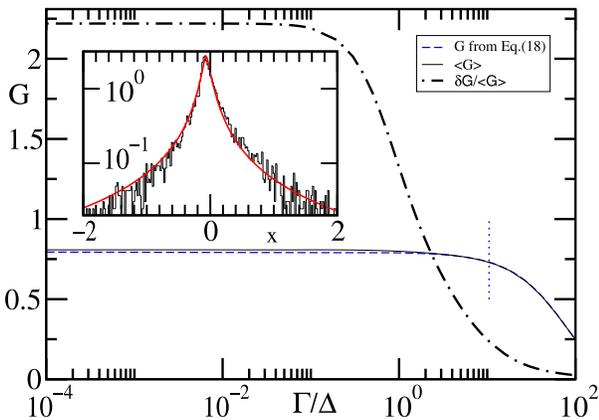,height=0.93\hsize, angle=-90,clip}
\caption{Comparison of the exact quantum result 
Eq.(\ref{e9}) for $\langle G\rangle$ with the diagonal approximation 
Eq.(\ref{e18}) in case of the network of Fig.1d. 
The average is taken over more than 20000 levels around $E_F$, 
while the calculation (for each Fermi level) was performed 
in an interval of 32000 levels. 
The valency $v$ of each vertex is picked up randomly. 
The transmission of the piston is $g_{0} \approx 0.1$.  
The perpendicular dotted line indicates the border of 
the regime where the QM calculation is valid (see text). 
We also plot the standard deviation $\delta G/\langle G\rangle$,  
while the inset displays the distribution $P(G{-}\langle G \rangle)$ 
for $\Gamma=0.0001\Delta$. Notice the slight asymmetry. 
The smooth line is the best fit with $0.068/(0.024+|x+0.065|^{1.5})$.
In these calculations the temperature is $T=0$.}
\end{figure}

If the condition $\Gamma \ll \Delta_b$ breaks down we enter into a {\em non-perturbative 
regime} where the QM recipe Eq.(\ref{e6}) does not hold. However, if 
the system has a {\em classical limit}, then it can be argued (on semiclassical grounds) 
that in the non-perturbative regime the {\em classical} calculation can be trusted. 
Since the classical calculation gives the {\em same} estimate as the diagonal approximation, 
it follows that there should be no apparent breakdown of validity as we cross from the 
perturbative to the non-perturbative regime. We have studied this issue \cite{rsp} 
in the context of energy absorption (the $G^{\tbox{ohm}}$ conductance).

It is now interesting to discuss what happens in the non generic case $b \sim 1$. For 
this purpose we consider two specific examples: a scatterer on a closed ring (Fig.1c);  
and a scatterer on a disconnected bond (Fig.1e but without the reservoirs).  In the first 
example the adiabatic result Eq.(\ref{e2}) can be recovered from Eq.~(\ref{e3}) by substituting
$g_T=g_0$, while in the second case $G=0$ because $g_T=0$. 
The latter result requires further discussion. 
Since $g_T=0$ the possibility of getting a steady state circulating 
current is blocked. The zero order adiabatic picture is as follows: 
At any moment the lowest energy levels are populated, 
hence at any moment the charge distribution 
is roughly uniform (ergodic). If we plot the current as 
a function of time, we find that the ``snow plow" dynamics 
is counter balanced by adiabatic passages of the particles through 
the moving barrier. The latter manifest themselves in the current 
as short spikes that compensate the otherwise steady (snow plow) 
flow of current. The statistical properties of the current 
should be regarded as the simplest example for the generalized 
universal conductance fluctuations that we have discussed previously. 
If the driving is non adiabatic, then the particles 
do not have the time to make adiabatic passages through the scatterer, 
and then the ``snow plow" dynamics becomes more effective. 
Thus the non-adiabatic translation of the scatterer induces 
a steady non-zero current in the bond, which is associated with 
accumulation of charge on the heading side of the scatterer  
and depletion in the trailing side. Hence the current within 
the bond becomes (during some transient period) 
of the same order of magnitude as in the case 
of an open system (Fig.~1e), where the reservoirs are assumed 
to be of an infinite size.


In summary, the Kubo approach to quantum pumping 
allows to explore the crossover from the strictly adiabatic 
``geometric magnetism" regime to the non-adiabatic 
regime. In particular we were able to derive specific results 
for the generalized conductance, using either 
classical stochastic modeling or diagonal approximation, 
which are supported by numerical analysis.

D.C. has the pleasure to thank M.~B{\"u}ttiker for 
an instructive visit in Geneve. This research
was supported by the Israel Science Foundation (grant No.11/02),
and by a grant from the GIF, the German-Israeli Foundation 
for Scientific Research and Development.

\onecolumngrid
\vspace*{-2mm}
\twocolumngrid


\end{document}